# Waveguide Components and Aperture Antennas with Frequency- and Time-Domain Selectivity Properties

M. Barbuto, *Senior Member, IEEE*, D. Lione, A. Monti, *Senior Member, IEEE,* S. Vellucci, *Student Member*, *IEEE*, F. Bilotti, *Fellow, IEEE,* and A. Toscano, *Senior Member, IEEE*

*Abstract*—Filtering modules are essential devices of modern microwave systems given their capability to improve the signal-to-noise ratio of the received signal or to eliminate the unwanted interferences. For discriminating between different components, a filter exhibits a frequency-selective response that, however, is not able to distinguish between different signals whose spectrum falls within the pass-band of the filter itself. In this regard, some electromagnetic structures exhibiting, at the same frequency, different responses depending on the waveform of the incoming waves have been recently proposed. In this paper, we extend the aforementioned approach to the case of a standard waveguide filtering module. In particular, by loading a band-pass filtering iris with a proper lumped-element circuit, we design a waveguide component able to distinguish between different pulsed waves, even at the same frequency, depending on their pulse width. Moreover, by using this filter for capping an open-ended rectangular waveguide, a radiating element with both frequency- and time-domain selectivity properties is presented. The structures discussed in this paper may pave the way to a new class of microwave systems that, being both frequency and time selective, are less sensitive to noise and interferences.

## I. INTRODUCTION

THE technological innovation underway for many decades on microwave components is stimulated by the continuous demand for increasingly efficient and reliable radio systems. Among the most widely used components of a microwave system, waveguide sections are typically used for transmitting electromagnetic energy from the source to the radiating element. In fact, these lines allow the propagation of electromagnetic waves in a guided structure with low losses. However, they have a metal bulky structure that, though ensuring high power handling, leads to considerable space occupancy. The same problem also affects waveguide feeds and horn antennas that, thus, cannot be used in microwave systems with strong space constrains. For this reason, many studies have been recently carried out in order to integrate, within a waveguide section or a horn antenna, other microwave elements able to manipulate the propagating signal in a desired way [1]- [8]. In this manner, although the waveguide or horn dimensions will be the same, the size of the overall microwave system will be minimized by reducing the needs for external elements. In particular, standard resonant structures, such as waveguide irises [9], can be inserted in a waveguide section for introducing a filtering behavior and, thus, attenuating the propagating signal for a specific frequency range. By properly acting on the irises shape, some further manipulations can be also implemented, such as linear-to-circular polarization transformation [2]-[3], polarization rotation [4], or polarization-based signal splitting [5]. Moreover, by loading the inclusions with non-linear elements, such as electronic diodes, filtering modules with power-dependent behavior have been conceived [10]-[11].

So far, the geometry, the mutual position and the loading circuit have allowed the manipulation of the propagating signals based on their polarization state, frequency content and power level. However, some absorbing and cloaking devices have been recently developed [12]-[15], which interact with the signal not only according to these characteristics but also depending on the time duration of the signal. According to these results, the waveform of an electromagnetic signal can be considered as a new degree of freedom, which can be exploited for numerous innovative applications. For instance, in [12], the "pulse width" of the impinging signal is used as a discriminating characteristic for absorbing or transmitting the signal itself, with possible applications in electromagnetic shielding. In [15], a waveform-dependent metasurface is conceived for hiding an antenna to a pulsed radar system while keeping its radiation characteristics unaffected for a continuous wave signal. In both these cases, the electromagnetic structure has been loaded with a proper lumped-element circuit, which affects the time-domain response of the overall system.

Inspired by these recent works, the main objective of this communication is to combine, on a single waveguide component, frequency- and time-domain selectivity properties. In particular, the idea is to load a standard band-pass filtering module with a proper circuit able to filter out the signals that do not exhibit the prescribed waveform, even if their spectrum falls within the pass-band of the filter itself. In other terms, the proposed innovative waveguide system allows the propagation of signals satisfying specific characteristics in both frequency

M. Barbuto, D. Lione, and A. Monti are with the Engineering Department, "Niccolò Cusano" University, Rome, 00166 Italy (e-mail: mirko.barbuto@unicusano.it, alessio.monti@unicusano.it).

S. Vellucci is with the ELEDIA Research Center (ELEDIA@UniTN - University of Trento), Trento, 038123 Italy (e-mail: stefano.vellucci@unitn.com).

F. Bilotti, and A. Toscano are with the Department of Engineering, "ROMA TRE" University, Rome, 00146 Italy (e-mail: filiberto.bilotti@uniroma3.it, alessandro.toscano@uniroma3.it).

This work has been developed in the frame of the activities of the research contract CYBER-PHYSICAL ELECTROMAGNETIC VISION: Context-Aware Electromagnetic Sensing and Smart Reaction, funded by the Italian Ministry of Education, University and Research as a PRIN 2017 project (protocol number 2017HZJXSZ).

and time domain.

The paper is organized as follows. In Section II, we present the design of a filtering module exhibiting both frequency- and time-domain selectivity properties. In Section III, we validate the effectiveness of our design by reporting the results obtained by performing some circuital-electromagnetic full-wave co-simulations. In Section IV, we use the designed module for capping an open-ended waveguide and, thus, obtain a radiating element with filtering properties in both frequency- and time-domain. Finally, in Section V we draw the conclusions.

## II. DESIGN OF THE FILTERING MODULE WITH BOTH FREQUENCY- AND TIME-DOMAIN SELECTIVITY

As well known, a simple method to create a band-pass filtering behavior in a standard waveguide is to insert, across the waveguide itself, a thin metal plate with one or more holes in it, *i.e.*, an iris [16]. By coupling several irises, the operating bandwidth of the overall filter can be controlled [17]. On the contrary, by acting on the iris shape, a high-pass, low-pass or band-pass filtering effect can be obtained [18]. As discussed in the Introduction, some other functionalities could be further added by acting on the iris shape. However, here, we are interested in adding a time-domain selectivity property to a waveguide filtering element. For the sake of simplicity, we start considering a standard waveguide iris consisting of an annular slot made on a thin metallic plate.

As shown in Fig. 1, the iris has been realized on a grounded dielectric substrate (RO$^{TM}$ 3003 with thickness of 4.2 *mm*) and has been placed inside a WR284 rectangular waveguide in order to fill completely its section and introduce a band-pass filtering behavior. In particular, as shown in Fig. 2, the geometrical and electromagnetic parameters of the filter, reported in Fig. 1, have been chosen to obtain a resonant frequency of 3 GHz.

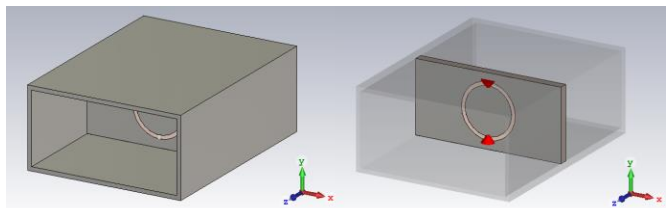

Fig. 1. Perspective view of the designed iris placed inside a standard WR284 rectangular waveguide section. The iris has been loaded with two discrete ports, which are required as communication points between the electromagnetic and the circuit simulators. The inner and outer diameters of the iris are 23.2 *mm* and 26.8 *mm*, respectively.

Like any other waveguide filter, the iris allows the propagation of an electromagnetic field that, in addition to respecting the boundary conditions imposed by the waveguide walls, is characterized by a frequency spectrum controlled by the filter response. The iris, thus, is not able to distinguish between two signals in the same frequency range, allowed by the filter response, even though their time-domain behaviors are different.

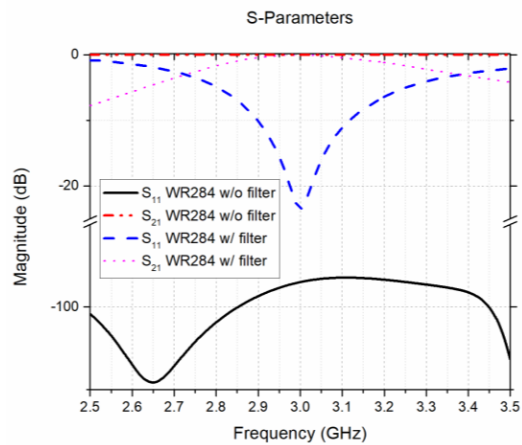

Fig. 2. Magnitude of the reflection and transmission coefficients of the WR284 waveguide section with and without the unloaded iris filter.

To add this functionality, we explore here the approach used in [12]-[15] for designing waveform dependent metasurfaces, which can distinguish between a short pulsed wave (PW) and a continuous wave (CW) signal.

In particular, the aforementioned approach is based on loading the metasurface with a proper lumped element circuit. Thus, in the commercial software CST Microwave Studio [19], we have loaded the iris with two "discrete ports", which allow integrating the response of a lumped-element circuit directly inside the full-wave simulator. In order to improve the interaction between the circuits and the electromagnetic wave, the discrete ports have been placed on the central vertical direction of the slot. Each discrete port connects the central metallic disk with the external part of the structure. In this way, as the propagating fundamental mode of the waveguide (the TE$_{10}$ mode) induces a potential difference between the two metallic parts of the iris, a current flows into the two circuits. The circuits, thus, are able to affect the electromagnetic response of the overall structure. In particular, assuming the signal spectrum lying in the pass-band of the filter, we could have two limiting cases. If a low-impedance circuit replace the discrete port, the current flows, almost unlimited, and the iris becomes a full metal plate that completely reflects the incoming signal. On the contrary, if a high-impedance circuit replaces the discrete ports, the iris is not affected by the circuit and the signal can propagate through the structure.

It is interesting to observe that, to switch between these two cases depending on the waveform or pulse width of the input signal, the loading circuits should exhibit a transient time-domain response, as first discussed in [12]. To do so, a simple solution relies on the use of a diode bridge and a few electronic passive elements like resistors (R), capacitors (C) and inductors (L). The diode bridge rectifies mostly to zero frequency the electromagnetic signal at its input port and the rectified signal is applied to the RLC elements, which respond differently for long and short pulses, even at the same input frequency. In the transient time-domain response of an RC parallel circuit, in fact, a strong current initially flows through the capacitor that behaves as a low-impedance element. However, the capacitor charges over time and, once it is fully charged up, it behaves like an open circuit.

Therefore, by loading the filtering iris with a parallel RC circuit we expect that short pulsed signals will be reflected, while continuous wave signals persisting over time will be transmitted through the screen. Conversely, in an RL series circuit, the current flow is initially counteracted by the inductor electromotive force, which decreases over time. In this case, the signal can pass through the screen on the first moments (when the circuit is in an open circuit condition) but the transmitted signal is increasingly attenuated over time. Moreover, by properly combining RC and RL circuits, band-pass or notched-band time-domain characteristics could be also obtained. Hence, in order to introduce a time-domain response in the iris, in our simulations we have considered four types of circuit configurations (reported in Fig. 3): an RC parallel circuit, an RL series circuit and the combination of these two circuits in parallel and in series.

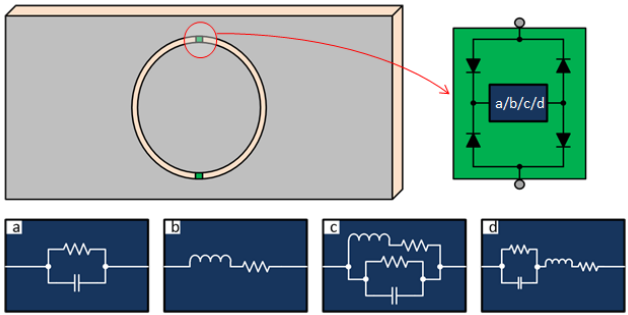

Fig. 3. The iris is loaded with some simple circuits composed by a diode bridge rectifier and: (a) a parallel RC circuit, (b) a series RL circuit, (c) a parallel RC circuit in parallel with a series RL circuit, or (d) a parallel RC circuit in series with a series RL circuit.

III. ANALYSIS OF THE RESULTS

In order to evaluate the effectiveness of the proposed structure to behave simultaneously as a filtering module in both frequency- and time-domain, we have performed some circuital-electromagnetic full wave co-simulations of the structure shown in Fig. 1, when loading the filtering iris with the different circuit configurations shown in Fig. 3. The simulation environment has been set for performing time-domain simulations and provide both the input and output signal trend over time and the transmittance of the system for an input pulsed signal at 3 GHz when varying its pulse width. In particular, the pulse width has been varied from 10 *ns* to 10 *us*. In addition, the results have been also extracted with different values of the reactive components (capacitor and inductors) while keeping the resistor value constant.

An important aspect to take into account during the simulations is the filtering bandwidth of the iris compared to the bandwidth of the input signal. In fact, a pulsed signal can be considered as the product between a phasor $sig(t) = A sin(\omega t + \phi)$ and a unitary impulse window:

$$\Pi_{T_{PW}}(t) = \begin{cases} 1, & t_0 < t < T_{PW} \\ 0, & t > T_{PW} \end{cases}$$

where $T_{PW}$ is the pulse width. Therefore, the signal spectrum is composed by the fundamental harmonic and many others minor harmonics due to the time window. The signal spectrum tends to be similar to the fundamental harmonic when the pulse width increase. By decreasing the signal pulse width, the secondary harmonics become more relevant until they become comparable to the fundamental one. Thus, a very short pulse in time is characterized by strong higher order harmonics spread over a large bandwidth, which can be out of the bandwidth of the filtering iris. In this case, the input signal can be affected by the frequency selectivity of the iris rather than by its time-domain behavior. Therefore, in order to avoid this phenomenon, it is necessary to define a lower limit for the pulse width, which, in our system, corresponds to a few nanoseconds.

The results of the circuital-electromagnetic full wave co-simulations are summarized in the next figures. In particular, Figs. 4 - 7 show the input and output signal amplitudes in the time-domain for the different circuit configurations. From these figures, we can see that the amplitude of the output signal is very different when varying the pulse width of the input signal preserving its fundamental frequency component.

In particular, in the RC case, a short signal ($T_{PW}$ = 10 *ns*) with a fundamental harmonic at 3 GHz is strongly attenuated due to the short-circuit condition introduced by the capacitor. However, as the pulse width increases ($T_{PW}$ = 1000 *ns*), the output signal tends to assume the same amplitude value of the input waveform. As discussed in the previous Section, this difference in the response of the filter is due to the time-domain behavior of the capacitor. In fact, if we consider that at $t = 0$ the capacitor is completely discharged, when the signal approaches the structure a current flows through the capacitor short-circuiting the edges of the iris. The latter, thus, can be considered as a full metal plate that fully reflects the impinging field over the entire frequency band. On the contrary, once the capacitor is charged up, only very low currents flow on the metallic iris, which, thus, can be considered as a standard band-pass frequency filter. Please note that when the signal is no longer present, the capacitor discharges on the resistor in parallel and, thus, the circuit is able to restore its behavior in case of another pulsed input signal.

In the dual case of an RL loading circuit, a short signal ($T_{PW}$ = 10 *ns*) with a fundamental harmonic at 3 GHz passes through the iris almost unchanged. However, as the pulse width increases ($T_{PW}$ = 1000 *ns*), the output signal tends to be progressively reflected by the iris. Again, the time-domain selectivity property is related to the reactive element of the circuit. In fact, at the beginning of the transient response, the inductor contrasts the flowing current with an opposing electromotive force (*fem*). Since the current flowing through the inductor is very low, the iris works as it were unloaded and the signal can propagate beyond the screen. However, when the inductor transient continues, the *fem* decreases over time and the current flowing through it increases. Therefore, for large pulse widths, the iris is short-circuited and behaves as a full metallic screen.

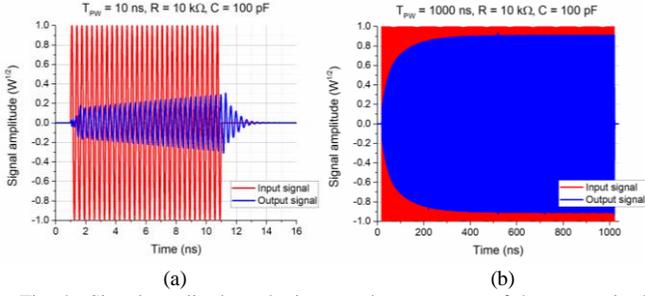

Fig. 4. Signal amplitude at the input and output ports of the system in the case of a parallel RC loading circuit for different pulse widths: (a) $T_{PW}$ = 10 ns, (b) $T_{PW}$ = 1000 ns. In the former case, the impinging signal is almost completely reflected though its frequency falls within the pass-band range of the filter.

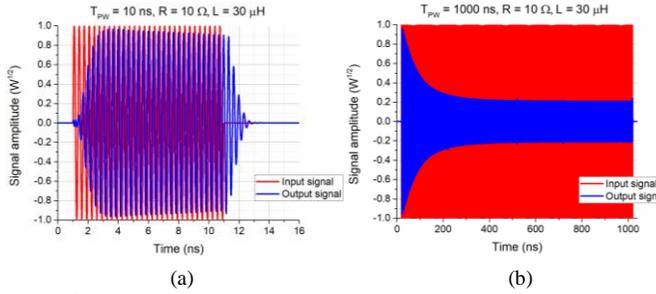

Fig. 5. Signal amplitude at the input and output ports of the system in the case of a series RL loading circuit for different pulse widths: (a) $T_{PW}$ = 10 ns, (b) $T_{PW}$ = 1000 ns. In the latter case, the impinging signal is almost completely reflected though its frequency falls within the pass-band range of the filter.

These results confirm that the time-dependent behavior of the reactive elements enables us to change the response of a microwave filter according to the pulse width of the incoming signal, even at the same frequency. In particular, a high-pass or low-pass behavior in the time domain can be obtained by simply using a rectifying diode bridge and a parallel RC circuit or a series RL circuit, respectively.

Fig.6 and Fig. 7 show the input and output signal waveforms for the circuit configurations in Fig. 3-c and Fig. 3-d. In these cases, due to the simultaneous presence of both inductive and capacitive elements, band-pass or band-stop time-domain selectivity can be obtained. In particular, using the parallel RC-RL configuration of Fig. 3-c, only a signal with a pulse width within the pass-band of the circuit can be transmitted. In the series RC-RL case of Fig. 3-d, the overall system exhibits instead a dual behavior, with a stop-band for specific pulse widths and the band-pass behavior in the frequency domain.

In order to confirm that time-domain selectivity property of the overall structure is due to the loading circuit altering the response of the iris, we have evaluated the magnitude of the loading impedance for different values of the pulse width of the incoming signal. As can be appreciated from Fig. 8, in the RC case, the impedance magnitude at 3 GHz tends to increase with the pulse width of the incoming signal, which is in agreement with the previous results. In fact, for a short signal with a fundamental harmonic at 3 GHz, the RC-loaded iris is short-circuited and the signal is almost reflected. On the contrary, as the pulse width increases, the loading impedance increases and the iris behaves as a standard band-pass frequency filter that allows the propagation of the signal at 3 GHz.

For RL loading circuit, a dual behavior is observed. In particular, the impedance magnitude tends to decrease with the pulse width of the incoming signal and, thus, an RL-loaded iris allows the propagation of short signals while reflecting signals with large pulse-width.

Finally, for parallel (series) RC-RL circuits, we have a maximum (minimum) of the impedance magnitude for intermediate values of the pulse-width, confirming the pass-band (notched-band) behavior in the time-domain of the overall structure.

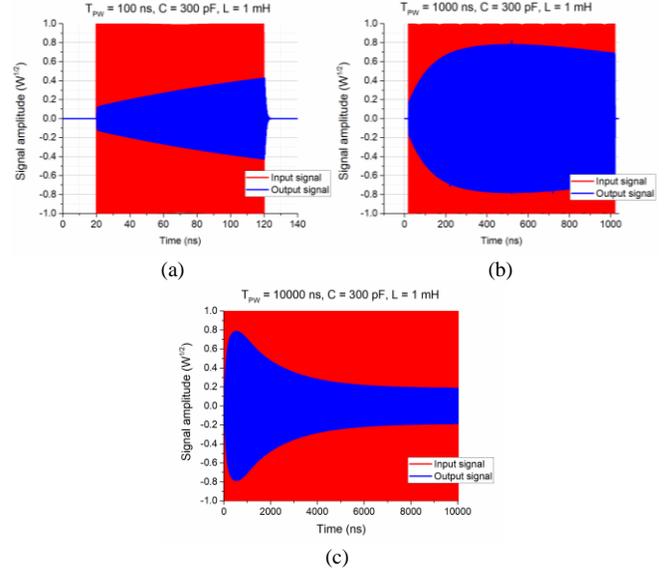

Fig. 6. Signal amplitude at the input and output ports of the system in the case of a parallel RC-RL loading circuit for different pulse widths: (a) $T_{PW}$ = 100 ns, (b) $T_{PW}$ = 1000 ns, (c) $T_{PW}$ = 10000 ns. In this configuration, the impinging signal is almost completely reflected except for the case in (b).

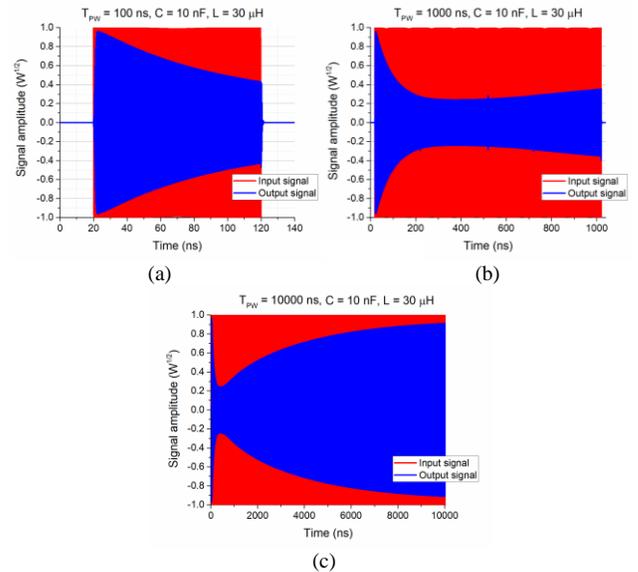

Fig. 7. Signal amplitude at the input and output ports of the system in the case of a series RC-RL loading circuit for different pulse widths: (a) $T_{PW}$ = 100 ns, (b) $T_{PW}$ = 1000 ns, (c) $T_{PW}$ = 10000 ns. In this configuration, the impinging signal is almost completely transmitted except for the case in (b).

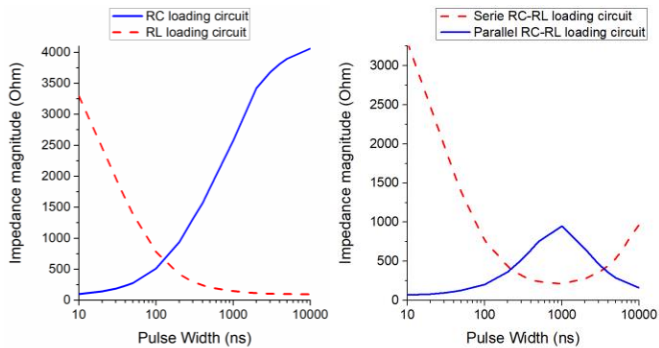

Fig. 8. Impedance magnitude at 3 GHz of: (left) the RC and RL loading circuit for different values of the pulse width; (right) the parallel and series RC-RL loading circuit for different values of the pulse width.

The time-domain filtering properties of all the four circuit configurations and the effect of the lumped elements values can be clearly inferred from Fig. 9 and Fig. 10. These figures show the transmittance, *i.e.* the ratio between the energy of the output and input signals calculated by entirely integrating transmitted and incident waveforms in the time domain [14], for different values of the reactive elements and as a function of the pulse width of the input signal (whose frequency is fixed at 3 GHz).

In particular, in the RC circuit case reported in Fig. 9-a, transmittance tends to increase with the pulse width of the incoming signal. The iris, thus, acts as a high-pass filter in the time-domain that allows propagation of only pulsed signals whose width is larger than a threshold value. The threshold can be controlled by simply acting on the capacitance value.

In the RL circuit case, whose results are reported in Fig. 9-b, the ratio between the output and input energy tends to decrease as longer the signal lasts over time. This corresponds to a low-pass filtering behavior that reflects all signals with a pulse width above a given threshold. Also in this case, the threshold can be controlled by acting on the reactive element of the circuit.

Finally, with the parallel or series combination of an RC and an RL circuit, a pass-band or notched-band temporal response can be obtained modifying the values of the reactive elements, as shown in Fig. 10-a and Fig. 10-b.

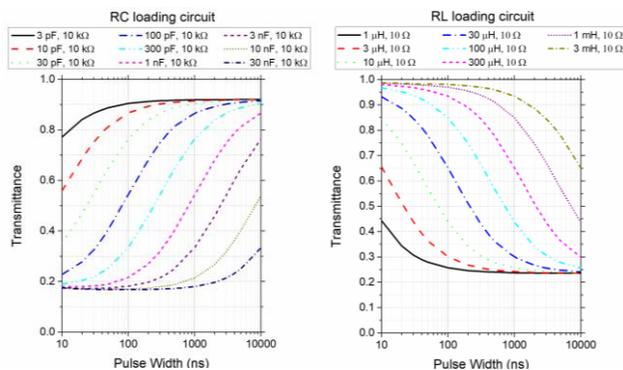

Fig. 9. Transmittance of the overall system for a parallel RC-based circuit (left) and a series RL-based circuit (right). Different values of the reactive elements are considered.

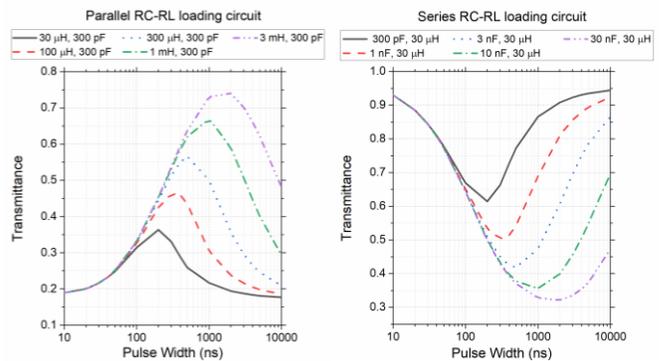

Fig. 10. Transmittance of the overall system for a parallel RC-RL circuit (left) and a series RC-RL circuit (right). Different values of the reactive elements are considered.

We remark here that all simulations have been carried out by taking into account the possible parasitic components of the diode and its realistic performances. In particular, we have used the spice model of the commercial diode MA4FCP200 [20]. The values of the lumped elements, instead, are reported in the legend of Fig. 9 and Fig. 10.

## IV. Design of a Waveguide Filtering Antenna with both Frequency- and Time-Domain Selectivity

Waveguide sections are not only used as closed transmission lines for transmitting signals between two distinct points, but are often left open for radiating in free-space. In particular, open-ended waveguides are typically used as feeding elements of parabolic reflectors or as receiving antennas for measuring system. In this case, as discussed in the introduction, filtering modules are often integrated in the overall structure for reducing the operating bandwidth of the antenna and improving the signal-to-noise ratio. In this scenario, by replacing the filtering module with the structure proposed in the previous Section, we can also conceive a new radiating structure that, in addition to filter the undesired spectral components out, is able to radiate/receive only electromagnetic signals with specific time-domain characteristics.

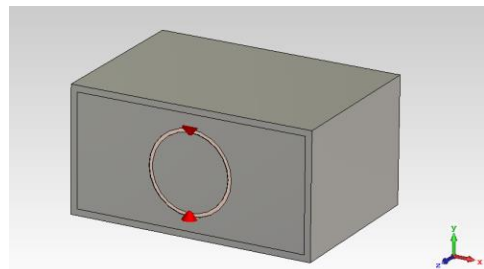

Fig. 11. Perspective view of the waveguide filtering antenna consisting of a WR284 waveguide section capped by the circuit-loaded iris. The inner and outer diameter of the iris are 23.9 mm and 26.1 mm, respectively.

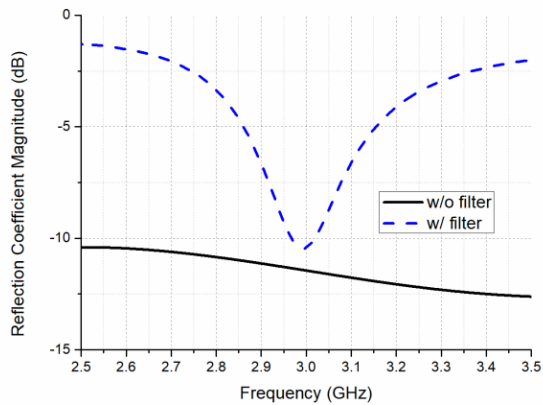

Fig. 12. Reflection coefficient magnitude of the WR284 waveguide section with and without the unloaded iris.

For validating this concept, we have simulated the response of the structure shown in Fig. 11 and consisting of a WR284 waveguide section capped by the previously reported circuit-loaded iris. In this case, we have slightly changed the iris dimensions, in order to obtain, as shown in Fig. 12, a good impedance matching only around the resonant frequency of 3 GHz. This behavior directly confirms the frequency-domain filtering properties of the antenna. In fact, the realized gain is directly related to the matching properties of the antennas and, thus, by narrowing its matching bandwidth, the radiation/reception performances of the antenna will be narrowed in frequency as well.

Through a proper set of circuital-electromagnetic co-simulations, we have analyzed the radiating properties when the iris is loaded with the RC-based diode bridge circuit of Fig. 3-a, with R = 10 $k\Omega$, C = 80 $pF$. In particular, we have reported in Fig. 13 and Fig. 14, the radiation patterns of the overall structure, in the two main planes, for signals with different pulse width. As can be appreciated, due to the time-domain behavior of the circuit, the antenna has poor radiation performances for short pulses ($T_{PW}$ < 100 ns), which short-circuit the filtering iris. On the contrary, increasing the pulse width of the signal transmitted/received by the antenna, the open-circuit condition of the iris is restored and the radiation performances resemble the ones of the unloaded filtering antenna.

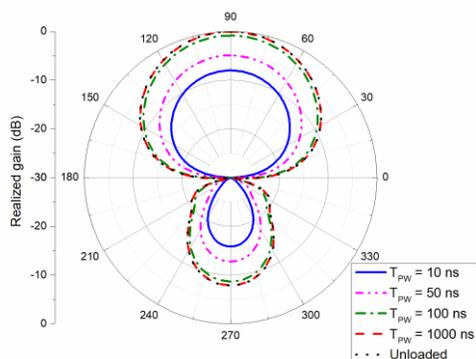

Fig. 13. Realized gain radiation diagram on the *xz*-plane of the waveform selective filtering antenna for different pulse width values at 3 GHz.

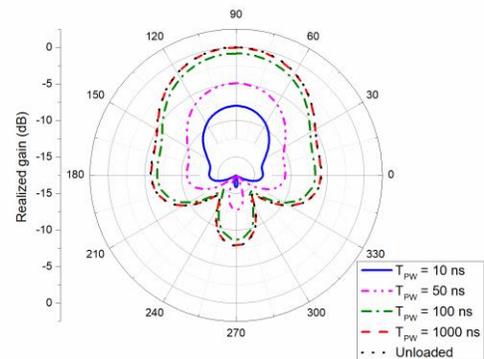

Fig. 14. Realized gain radiation diagram on the *yz*-plane of the waveform selective filtering antenna for different pulse width values at 3 GHz.

Please note that, for the sake of brevity, we have reported here only the case of a RC-based loading circuit. However, similar results can be obtained with all the circuits reported in Fig. 3. Therefore, by properly choosing the circuit configuration, we can design a band-pass filtering antenna in the frequency domain that also exhibits a high-pass, low-pass, band-pass or notched-band behavior in the time domain.

Finally, we would remark here that simulations have been carried out at fixed power of 1 W (corresponding, thus, to a more realistic scenario in transmission mode). In this way, all the diodes are turned on by the propagating signal that, thus, is actually rectified by the diode bridge. In the receiving mode, we expect that the performance of the proposed antenna could be affected by the power-level of the received signals that, in some cases, could be not enough for turning on the diodes. However, a power-dependent analysis of the system is beyond the scope of this work and could be the subject of further research.

## V. Conclusion

To summarize, in this communication we have designed a waveguide filtering module that, in addition to filter out-of-band noise or interfering signals, is able to discriminate between different waveforms, even at the same frequency, depending on the pulse width of the signals. For this purpose, we have integrated in the structure of a standard waveguide iris a lumped-element non-linear circuit able to convert almost to zero frequency the impinging electromagnetic wave and exhibit a waveform-dependent response. Then, the proposed filter has been used to cap an open-ended waveguide antenna and, thus, to design a radiating element that exhibits both frequency- and time-domain selectivity properties. The effectiveness of the aforementioned structures has been validated by performing circuit-electromagnetic co-simulations and taking the realistic response of a commercial diode into account.

We remark here that the proposed structure may pave the way to a new class of radiating systems that, being specifically designed for working with a given waveform and in a given bandwidth, are more robust to interference from the external environment.